\documentclass[submission,copyright,creativecommons]{eptcs}
\providecommand{\event}{AREA workshop} 

\usepackage{iftex}
\usepackage{graphicx}
\usepackage{subcaption}
\usepackage{latexsym}
\usepackage{xcolor}
\usepackage{soul}
\usepackage{array}
\usepackage{mathtools}
\usepackage{multirow}
\usepackage{booktabs} 
\usepackage{moresize}
\usepackage{amsmath}
\usepackage{amssymb}
\usepackage{xcolor,colortbl}
\usepackage{algorithm}
\usepackage[noend]{algpseudocode}
\usepackage{underscore}
\usepackage[english]{babel}
\usepackage{listings}
\usepackage{nccmath}
\ifpdf
  \usepackage{underscore}         
  \usepackage[T1]{fontenc}        
\else
  \usepackage{breakurl}           
\fi

\title{Online Proactive Multi-Task Assignment with Resource  Availability Anticipation }
\author{Déborah Conforto Nedelmann \quad   Jérôme Lacan \quad  Caroline Chanel
\institute{ISAE-SUPAERO, Université de Toulouse, France}
\email{\{deborah.conforto-nedelmann,jerome.lacan,caroline.chanel\}@isae-supaero.fr}
}
\def\titlerunning{Online Proactive Multi-Task Assignment with Resource Availability Anticipation }
\def\authorrunning{D. Conforto Nedelmann, J. Lacan \& C. Chanel}
\begin{document}
\maketitle
\vspace{-4mm}
\begin{abstract}
With the emergence of services and online applications as taxi dispatching, crowdsourcing, package or food delivery, industrials and researchers are paying attention to the online multi-task assignment optimization field to quickly and efficiently met demands. In this context, this paper is interested in the multi-task assignment problem where multiple requests (e.g. tasks) arrive over time and must be dynamically matched to (mobile) agents. This optimization problem is known to be NP-hard. 
In order to treat this problem with a proactive mindset, we propose to use a receding-horizon approach to determine which resources (e.g. taxis, mobile agents, drones, robots) would be available within this (possibly dynamic) receding-horizon to meet the current set of requests (i.e. tasks) as good as possible. Contrarily to several works in this domain,  we have chosen to make no assumption concerning future locations of requests. To achieve fast optimized online solutions in terms of costs and amount of allocated tasks, we have designed a genetic algorithm based on a fitness function integrating the traveled distance and the age of the requests.
We compared our proactive multi-task assignment with resource availability anticipation approach with a classical reactive approach. The results obtained in two benchmark problems, one synthetic and another based on real data, show that our resource availability anticipation method can achieve better results in terms of costs (e.g. traveled distance) and amount of allocated tasks than reactive approaches while decreasing resources idle time.
\end{abstract}
\vspace{-4mm}
\section{Introduction}
\vspace{-2mm}
The emergence of new applications dedicated to services such as taxi dispatching \cite{Dickerson_2018}, ridesharing \cite{Herbawi_2011}, crowdsourcing \cite{Wang_Yang_2020} or package delivery has generated a lot of interest in the field of online multi-task assignment. What all these services have in common, is that users can make requests and the platforms have to adapt their resources in order to satisfy the demand. 

The difference between offline and online assignment is that on the online one, requests (i.e. tasks) arrive over time and are dynamically matched whereas, in the offline case, all the requests are known beforehand \cite{Dickerson_2018}. With offline matching, finding an optimal task assignment is doable whereas online matching poses additional challenges. In a majority of online task assignments, resources are disposable and can only be used once, whereas we will consider the case where the resources are reusable in accordance with the following references \cite{Sumita_Ito_Takemura_Hatano_Fukunaga_Kakimura_Kawarabayashi_2022}, \cite{Dickerson_2018}. 
In other words, after an agent has received an assignment, it will not be available for a new allocation for a certain period of time before being able to be assigned to new requests. 

In order to treat online (multi-)task assignment, there are generally two options to manage the time. The first option consists to divide the time into intervals. At the end of each interval, the requests that have arrived within this interval are assigned to agents that have finished their previous tasks \cite{Alonso_Mora_2017}, \cite{Lesmana_2019}. The second option consists in assigning the requests immediately upon their arrival (i.e. continuously on time) to whatever available agents at that moment. Both of these solutions have drawbacks: the first one might result in a better solution but the waiting time before agents are available to assign tasks might be important and the second one assigns the request immediately but the solution is worse \cite{Sumita_Ito_Takemura_Hatano_Fukunaga_Kakimura_Kawarabayashi_2022}.

Following the literature, there are several possibilities to describe the requests according to their arrival model \cite{Mehta_2012} but, in this work, we have chosen to make no assumption on any arrival distribution. This choice was motivated by the fact that we chose to focus on the performance of online task allocation algorithms without being distracted by any knowledge of future task appearances. 

With a proactive mindset, our contribution to treat this problem is to use a receding-horizon approach to determine which agents would be available at a given time horizon given their current allocated tasks. Inspired by \cite{Alonso_Mora_2017} and \cite{Lesmana_2019}, our approach divides the time in small intervals and accounts for a certain horizon size (some time steps in the future), to determine where the agents would be and if they would have finished their tasks within this horizon. This would lead to the choice to consider them as near-future available resources for current multi-task assignment problem. Hence, our contribution can be seen as a  proactive multi-task assignment with resource availability anticipation approach. 

In terms of assignment computations, this problem is similar to the Multiple Traveling Salesmen Problem with open path and multiple depot \cite{Patel_2020}. A  way to treat this problem online is to use meta-heuristic approaches, such as Genetic Algorithm, as we will present in next section. Moreover, meta-heuristic approaches allows to use multi-objective criteria costs to evaluate (multi-)task assignments solutions \cite{Rangriz_2019}. 

We compared our proactive approach with a classical reactive approach (no anticipation) using two benchmark problems: a synthetic and a real-data based one. In brief, the results show that our resource availability anticipation method can achieve better results in terms of costs (e.g. traveled distance) and amount of allocated tasks than a reactive approach. Interestingly, the results also demonstrate that our approach decreases resources idle time.

The paper is organized as follows. Section \ref{sec:related} overviews related works. Section \ref{sec:prob} formally describes the optimization problem we are addressing. The approach proposed to treat the proactive online multi-task assignment problem is presented in Section \ref{sec:approach}. Experiments are showed in Section \ref{sec:expes}, and Section \ref{sec:conclusion} concludes the paper.

\vspace{-4mm}
\section{Related Work}
\vspace{-3mm}
\label{sec:related}
The general problem we address in this paper is the multi-task assignment with reusable resources, where the goal is to coordinate a set of agents (i.e. the resources) in order to accomplish some tasks in an efficient way \cite{Khamis_2015} within a given time period. Here, we focus on the case where multiple agents have to travel, reach and perform their tasks in a way that the overall (traveled) distance is minimized and the number of assigned tasks is maximized. This is reminiscent of the Multiple Traveling Salesmen Problem (MTSP), a well known hard optimization problem, for which literature has proposed several solution search algorithms \cite{Patel_2020}\cite{Khoufi_MTSP_2021}. In particular, our problem is similar to the special case of MTSP with open path and multiple depots. For this last, most common methods use meta-heuristics which include Ant Colony Optimization, Simulated Annealing and Genetic Algorithm \cite{Khoufi_MTSP_2021}\cite{Patel_2020}. 

In this paper we are focusing on online variants of the classical multi-task assignment approaches \cite{Khamis_2015}, \cite{Hussein_2013}, \cite{Choi_2009}, where requests arrive dynamically. Compared to classical offline models, online models tend to perform worse due to the uncertain nature of the future requests \cite{Buchbinder_2007}.
In spite of this, several methods have been used to find solutions for the online assignment problem. A common one is the Linear Programming \cite{Sumita_Ito_Takemura_Hatano_Fukunaga_Kakimura_Kawarabayashi_2022} \cite{Nanda_2020}. Evolutionary algorithm have also been used due to their strong performance for multi-criteria objective \cite{Rangriz_2019}. Given the nature of the problem we are treating in this paper, this solving approach interests us as we will detail on Section \ref{sec:approach}. Reference \cite{Herbawi_2011} has studied, in the context of online assignment, the performance of variants of the Ant Colony Optimization (ACO) algorithm compared to a Genetic Algorithm (GA) and have found that a variant of ACO would perform at best the same as the GA. This is why GA seems to be an interesting choice.

Additionally, we have observed in the literature  (see \cite{Alaei_2012} \cite{Gong_2022} \cite{Dickerson_2018} \cite{Sumita_Ito_Takemura_Hatano_Fukunaga_Kakimura_Kawarabayashi_2022} \cite{Hikima_2022}) that papers tend to focus on using a request arrival model, usually a distribution that describes the likely tasks locations (e.g. requests). Several authors have examined the concept of anticipation by working with the distribution. In \cite{Burns_2012}, the authors divides the time in intervals and with the arrival distribution determines the number of tasks and their location a couple of intervals ahead, enabling them to better position their agents to respond to the future tasks more effectively. In \cite{Filippo_2019}, the authors have beforehand an example of the characteristics of the requests which allows them to build offline a distribution about the more likely scenarios. On a more practical example, both \cite{Schreckenberf_2001} and \cite{Claes_2011} chose to build an anticipation model to try to avoid traffic jams thanks to previous registered data. As said previously, we make no assumption about the requests arrival model, but we work on anticipating the availability of agents given their current tasks (i.e. the end-time of the tasks assigned to a given agent and its corresponding location).  As far as we know, there is no work in the literature for online proactive multi-task assignment with resource availability anticipation. 

Moreover, to treat the problem online, time has to either be divided in small intervals or requests must be assigned immediately at their arrival. Both have their benefits and drawbacks, because as presented in \cite{Wang_Bei_2022}, either the new allocation is sub-optimal (regarding a long-term horizon), due to a lack of available agents at a given moment compared to the fleet, or that, in order to find a better solution, we need to wait for more agents to become available, which can be long. In this work, we have decided to divide the time into intervals and assign the requests that have arrived within the interval to agents that may be available within a given receding-horizon. 

\vspace{-4mm}
\section{Problem Statement and Treatment}
\label{sec:prob}
\vspace{-3mm}
\subsection{Problem statement}
\vspace{-3mm}
Multi-task assignment is a combinatorial optimization problem. In our case and over the entire time horizon, we aim to maximize the amount of allocated requests (i.e. tasks) while also minimizing the overall traveled distance of the agents (i.e. the resources). 

We consider that we have a set of \textit{M} agents denoted by $A = \{a_1, a_2, …, a_M\}$. The location of the agents is denoted by $\mathbf{p_a}$ for all $a \in A$. The tasks are described by their location $\mathbf{p_r}$ and the expected time $t_r$ an agent is supposed to met them. Thus, a single request $r \in R$ can be written as $r = (\mathbf{p_r},  t_r)$. A request can be assigned to only one agent. Thus, each agent $a\in A$ has an associated vector of requests assigned to them: $R_{a} = (r_a^1, r_a^2, …, r_a^n)$, where \textit{n}  is the number of tasks assigned to $a$ at a given time. To describe if a request has been allocated to an agent, we use the binary variable $x_{a,r}$. If request $r$ was allocated to agent $a$, we have $x_{a,r} = 1$, otherwise $x_{a,r} = 0$. Thus, we have $R_a = \{\bigcup_{r \in R} r | x_{a,r}=1\}$ defining the set of tasks from $R$ assigned to $a$. To an agent a maximum of $C_{a}$ requests can be assigned. Let $L_{a}$ be the cost associated to the path length that an agent have to travel to accomplish their assigned tasks at a given time.

The total time horizon $T$ is divided into small intervals called time steps or time windows. The time windows are indexed by $\tau$ and their duration is a constant equal to  $\delta$.  During the time window $\tau$, we read the buffer of requests \textit{B} and store the new requests in $R_\tau$. It is worth saying that $R_a$, $C_a$ and $L_a$ may also depend on this time step $\tau$, such as $R_{\tau,a}$, $C_{\tau,a}$ and $L_{\tau,a}$.


Therefore, the general optimization problem we address can be formalized as:
\begin{equation} \medmath {
        \min \left [ \sum_{\tau =0}^{T-1} \Big ( \alpha \sum_{a \in A} L_{\tau,a}  \big ( \bigcup_{r_i \in R_\tau} r_i | x_{a,r_i}=1 \big )  + (1-\alpha) \big ( \lvert R_\tau \rvert - \sum_{a \in A} \sum_{r \in R_\tau} x_{a,r}\big )\Big ) \right ]}
\end{equation}
\leftline{subject to:}
\begin{equation} \medmath{
   \sum_{a \in A} x_{a,r} \leq 1, \forall r \in R_\tau,}
\end{equation}
\begin{equation} \medmath{
    \sum_{r \in R_\tau} x_{a,r} \leq C_{\tau,a}, \forall a,  \forall \tau \in \{0, ..., T-1\}}
\end{equation}
where the weights $\alpha (\in  [0,1])$ and $(1-\alpha)$ define the relative importance of the total path length cost and the total number of assigned tasks, including criteria scaling needs. 
We do not have any information on the upcoming requests, thus the general optimization problem formalized above is hard (or even impossible) to be solved offline.
\vspace{-4mm}
\subsection{Problem treatment}
\vspace{-3mm}
With a proactive mindset, we propose to use a receding-horizon approach to determine which agent would be available at horizon $H$ given their current allocation.  For simplicity, we define this receding-horizon as a multiple of the duration of the time intervals  such as, $H(k) = k \delta$, with $k\geq 0$ (e.g $H(5) = 5\delta$).
We will call the set of agents that are available within horizon $H$ at time step $\tau$ as $A_\tau(H) \in A$. Note that the number of available agents depends on the size of the receding-horizon $H$ and the time step $\tau$.

So to achieve a solution to the general optimization problem presented, we use this receding-horizon to be proactive, and we adapt the optimization problem to be solved at each time window $\tau$. At $\tau$, we know the position of the tasks $R_\tau$ and we can check the availability of the agents within this receding-horizon $H$.

Thus, at the time step $\tau$, the optimization problem we solve is the following:
\begin{equation} \medmath{
    \min \left [  \alpha \sum_{a \in A_{\tau}(H)} L_{\tau,a}  \Big ( \bigcup_{r_i \in R_\tau} r_i | x_{a,r_i}=1 \Big ) \Big . \right . + \Big . \left .  (1-\alpha) \Big ( \lvert R_\tau \rvert - \sum_{a \in A_{\tau}(H)} \sum_{r \in R_\tau} x_{a,r}\Big ) \right ]}
\end{equation}
\vspace{-2mm}
\leftline{subject to:}
\vspace{-2mm}
\begin{equation} \medmath{
    \sum_{a \in A_\tau(H)} x_{a,r} \leq 1, \forall r \in R_\tau}
\end{equation}
\vspace{-2mm}
\begin{equation} \medmath{
    \sum_{r \in R_\tau} x_{a,r} \leq C_{\tau, a}, \forall a \in A_\tau(H)}
\end{equation}

In the following, we define how we compute $L_{\tau,a}$, the cost associated to the path length that an agent $a$ has to travel to accomplish its assigned tasks at $\tau$: 
\begin{equation} \medmath{
    L_{\tau,a} (\bigcup_{r_i})= c(\mathbf{p_a}; \mathbf{p_{r_1}})+ \sum_{i = 1}^{\lvert \bigcup_{r_i}\rvert -1} c( \mathbf{p_{r_i}}; \mathbf{p_{r_{i+1}}}), \forall r_i \in R_\tau}
\end{equation}

\noindent where $c$ is the cost associated to the distance between two locations and $\bigcup_{r_i} = \lbrace \bigcup_{r_i \in R_\tau} r_i | x_{a,r_i}=1 \rbrace$. This path cost formula was inspired by the definition of cost for MTSP with multi-depot open path proposed in \cite{Khoufi_MTSP_2021} and adapted to our case. $L_{\tau, a} (\bigcup_{r_i})$ effectively quantifies the length of the shortest path between the agent's location and his assigned tasks.

As a feature, and with the aim of maximizing the number of achieved requests within the entire horizon $T$, at $\tau$, we consider the tasks that have not been allocated in earlier time steps in addition to the current buffer $B$ to form $R_\tau$. Earlier requests are considered with higher priority compared to later tasks. This constraint will be integrated into the design of the solution, explained in Section \ref{sec:sol_init}. The reason we take into account this aspect is that we want to avoid cases where some tasks would potentially wait for a long time before being allocated or even never be allocated. 
\vspace{-3mm}
\section{Online Proactive Task-Assignment with Resource Availability Anticipation}
\label{sec:approach}
\vspace{-2mm}
\subsection{General solving procedure}
\vspace{-2mm}
In accordance with our problem statement and model stated previously, we summarize our general algorithm as the pseudo-code in Algorithm 1. In line 1 and 2, we first define the set of agents and their locations as well as the size of our receding-horizon $H$. 
At each time step $\tau$, we consider the tasks that have arrived since the last assignment which are stored in buffer $B$ along with the ones that have not been assigned yet (line 5). Then in line 6, we check the availability of agents at the receding-horizon $H$. Agents that are available within $H$ are then used for multi-task assignment using a genetic algorithm in line 7 (explained later). 
\begin{center}
\begin{minipage}{.48\linewidth}
\begin{algorithm}[H]
\caption{General Algorithm}\label{gen_algo}
\begin{algorithmic}[1]
\footnotesize
\State Definition of the set of agents $A$ and their locations
\State Definition of horizon $H$
\State $R_{\tau=0} \gets \emptyset$, $t_r=0$
\For{each time step $\tau$} 
\State $R_\tau \gets \mathrm{GetTasksFrom}(B,R_{\tau -1})$
\State $A_\tau(H) = \mathrm{AvailabilityAnticipation}(H, A, \tau)$
\State $\mathrm{Sol}_\tau = \mathrm{GeneticAlgorithm}(R_\tau, A_\tau(H), C_{\tau,a})$
\EndFor
\end{algorithmic}
\end{algorithm}
\end{minipage}
\hskip 7pt
\begin{minipage}{.48\linewidth}
\begin{algorithm}[H]
\caption{Availability Anticipation Process}\label{euclid}
\begin{algorithmic}[1]
\footnotesize
\Procedure{AvailabilityAnticipation}{H,A,$\tau$}
\State $\textit{$A_\tau(H)$} \gets \text{[]}$
\For{$\textit{$a \in A$}$}
\State $R_{\tau,a} \gets \bigcup_{r_i \in R_\tau} r_i | x_{a,r_i}=1$
\State $t_{R_{\tau,a}} \gets \mathrm{getLastTaskElementTime}(R_{\tau,a})$.
\If {$t_{R_{\tau,a}} < \tau + H$}
\State append $a$ to $A_\tau(H)$
\EndIf
\EndFor
\State return $A_\tau(H)$.
\EndProcedure
\State \textbf{end procedure}
\end{algorithmic}
\end{algorithm}
\end{minipage}
\end{center}

\vspace{-4mm}
\subsection{Availability Anticipation}
\vspace{-2mm}

We now define the essential part of our contribution which corresponds to the function on line 6 in Algorithm 1. The main difference with other reactive matching algorithms \cite{Sumita_Ito_Takemura_Hatano_Fukunaga_Kakimura_Kawarabayashi_2022} \cite{Hikima_2022} \cite{Dickerson_2018} is we are not using only the agents that are strictly available at $\tau$, but we are computing which agents will be available within a certain horizon. 

Using only the instantly available agents at the current time may provide sub-optimal solutions in terms of distance and the overall amount of requests allocated \cite{Wang_Bei_2022}. Note if an agent becomes available right after the time window, it will have to wait almost the time window duration before being assigned to new requests. We believe anticipation may allow to find better solutions in terms of overall traveled distance while being proactive in terms of resource usage.
Additionally, in terms of the number of requests completed, if we have more agents available, the number of tasks that can be allocated is automatically higher given an agent can only treat a maximum of $C_{\tau,a}$ requests per round. Another point is we do not wait for the agents to complete their previous tasks before assigning them new ones. With this, we avoid the problem of agents staying idle for almost a round. 

A feature of our solution is to try to avoid unbalance of tasks between agents where the requests of busy agents could be completed well after some newer requests, which are assigned to less busy agents.  With our solution, if some agents are very busy, they will simply not be picked for a new allocation until they complete some of their previously assigned requests. 

The availability anticipation process is detailed in Algorithm 2. For every agent, we check the estimated time $t_{R_{\tau,a}}$ when their last request is supposed to be completed in lines 4 and 5. If this time is within the horizon $H$ with respect to current time $\tau$ (line 6), then we consider this agent as available at the receding-horizon and consequently this agent is used for multi-task assignment. 
The last request completion time $t_{R_{\tau,a}}$ can be predicted by supposing each agent has a constant velocity. Using the current set of tasks assigned to an agent, $R_{\tau,a}$ we can calculate the time necessary for the agent $a$ to go to them and, also its final location.

\begin{algorithm}[!b]
\caption{Genetic Algorithm}\label{euclid}
\begin{algorithmic}[1]
\footnotesize
\Procedure{GeneticAlgorithm}{$R_\tau$, $A_\tau(H)$, $C_{\tau,a}$  }
\State Definition of the probability of mutation $p_{muta}$ and the probability of swapping $p_{swap}$
\State Definition of the size of the population $n_{Pop}$ and the size of a chromosome: $|A_\tau(H)| \times C_{\tau,a}$ 
\State Definition of the timing beginning $t_{beg}$
\State Fill each chromosome $chr$ of the population $Pop$ with $-1$
\State $\textit{$P$} \gets \text{[]}$
\For {$\textit{$r_i \in R_\tau$}$}
\State Calculate $p_i$ with the Boltzmann Probability Distribution
\State Append $p_i$ to $P$
\EndFor
\For {each $chr \in Pop$}
\While {$R_\tau \neq \emptyset$ \textbf{or} $count(v$ in $chr$ | $v= -1) \ne 0$}
\State Select $r_i$ from $R_\tau$ using $P$
\State Select $pos$ position in $chr$ randomly
\State $chr[pos] = r_{i_{ID}}$ 
\State Remove $r_i$ from $R_\tau$ 
\EndWhile
\EndFor
\State Set $min_{now} = \infty$ and $min_{prec} = 0$
\State Definition of the current time $t_{now}$
\While{$t_{now} - t_{beg} < \delta$ or $\lvert min_{now} - min_{prec} \rvert > \epsilon$}
\State $\textit{$S$} \gets \text{[]}$
\For{each $chr$ in $Pop$}
\State $s = fitness(chr)$
\State Append $s$ to $S$
\EndFor
\State Set $min_{prec} = min_{now}$ and $min_{now} = min(S)$
\State Select the best $30\%$ chromosomes $Best_{chr} \subset Pop$ regarding $S$ 
\State $Pop \gets \text{[]}$
\State Append $Best_{chr}$ to $Pop$
\While{$\lvert Pop \rvert < n_{Pop}$}
\State $parent1 = random(Best_{chr})$ and $parent2 = random(Best_{chr})$
\State $child_{chr} = Crossover(parent1, parent2)$
\State $v = random(0,1)$
\If{$v < p_{muta}$ and $v < p_{swap}$}
\State $child_{chr} = Swapping(child_{chr})$
\EndIf
\If{$v < p_{muta}$ and $v > p_{swap}$}
\State $child_{chr} = Inversion(child_{chr})$
\EndIf
\State Append $child_{chr}$ to $Pop$
\EndWhile
\State Update the current time $t_{now}$
\EndWhile
\State Set Sol$_\tau = chr^*$, such as $chr^* = argmin_{chr \in Pop} S$
\State \textbf{return} Sol$_\tau$
\EndProcedure
\State \textbf{end procedure}
\end{algorithmic}
\end{algorithm}

\vspace{-4mm}
\subsection{Genetic Algorithm}
\vspace{-2mm}
The Genetic Algorithm (GA) is an evolutionary algorithm and a well known meta-heuristic approach \cite{Holland_GA_1975} usually used to solve high combinatorial optimization problems \cite{Khoufi_MTSP_2021}, which iteratively constructs solutions. A proposition of a solution by a GA is called a chromosome and a list of chromosomes is a population. At each iteration, the population is evaluated and the better chromosomes are used to create a new generation thanks to some operations inspired by natural evolution theories. The new generation will hopefully provide better solutions than the previous one. This process is repeated until some stopping condition is reached. 

In the following, we detail some design choices of the GA that we use in order to find assignment solutions to the $R_\tau$ tasks considering the $A_\tau(H)$ agents. This operation corresponds to line 7 of Algorithm 1 and is detailed in Algorithm 3.

\vspace{-4mm}
\subsubsection{Solution representation}
\begin{figure}[!hb]
\centering
\begin{minipage}{0.48\linewidth}
\includegraphics[width=1\textwidth]{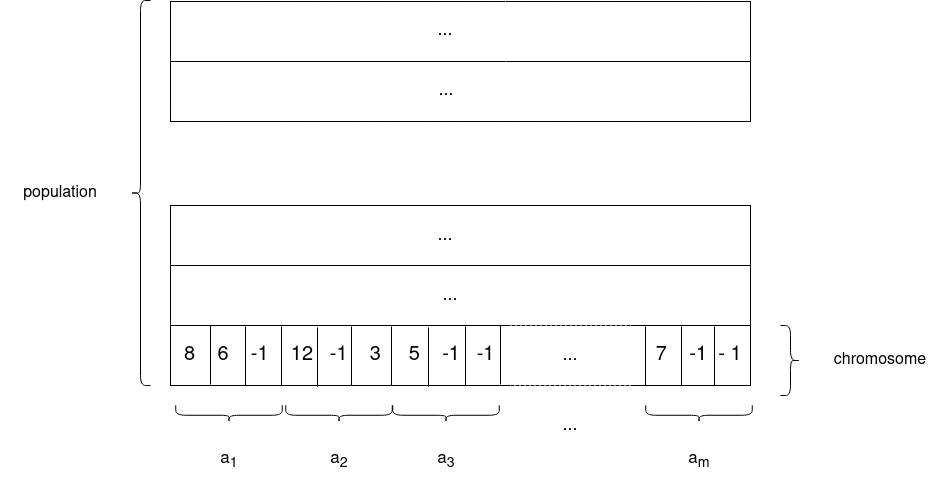}
\caption{Structure of the population and chromosome.} \label{fig:pop}
\end{minipage} \hfill
\begin{minipage}{0.48\linewidth}
\centering
\includegraphics[width=1\textwidth]{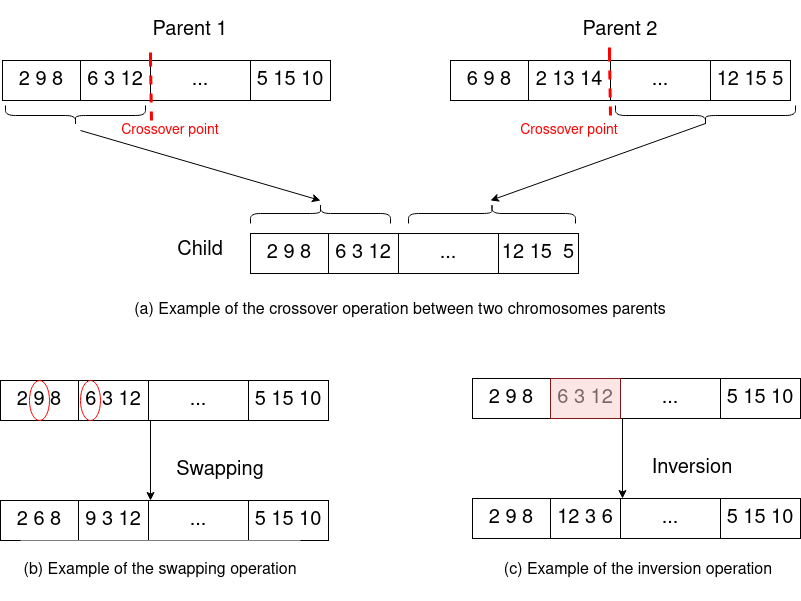}
\caption{Illustration of the different evolution operations: (a) for crossover and (b)  and (c) for the respective swapping and inversion mutation operations.}
\label{fig:operations}
\end{minipage}
\end{figure}

\vspace{-2mm}
Our population has the general aspect illustrated in Fig. \ref{fig:pop}. Each chromosome represents a proposition of solution (i.e. an multi-task assignment solution given the agents and requests considered). The size of a chromosome is therefore limited to $|A_\tau(H)| \times C_{\tau,a}$ to respect the maximum number of requests assigned to each agent. If this constraint is relaxed, the size is then $|A_\tau(H)|\times|R_\tau|$.
Each request has a unique positive identification number (ID) which is used inside the chromosome. If an agent is assigned less than $C_{\tau,a}$ tasks, the remaining values are noted as $-1$ to differentiate from the ID. For example, in the last chromosome of Figure \ref{fig:pop}, the GA has assigned requests $8$ and $6$ to the first agent, requests $12$ and $3$ to the second agent, request $5$ to the third agent, and request $7$ to the last agent. 
\vspace{-4mm}
\vspace{-4mm}

\subsubsection{Solution Initialization}
\vspace{-2mm}
\label{sec:sol_init}
This step corresponds from line 5 to 22  of Alg. 3. In line 5, we fill each chromosome with the value -1 that we will possibly replace with the ID of the requests. As mentioned previously, we want to prioritize the selection of tasks that have been registered earlier. For this, we use the Boltzmann Probability Distribution, which was proposed for thermodynamic field but has since been picked up to be used in the domain of reinforcement learning \cite{Boltzmann}. We propose the following adaptation:
\begin{equation} \medmath {
    p_i = 
    \begin{cases}
    \frac{1}{Q} \exp( -\frac{t_{r_i}}{\tau})& \mathrm{for}\;  \tau>0 \\
    \frac{1}{|R_{\tau}|} & \mathrm{for}\; \tau=0
    \end{cases}}
\end{equation}
\noindent where $Q = \sum_{r_j=1}^{|R_\tau|} \exp( -\frac{t_{r_j}}{\tau})$. Then, $p_i$ is the probability of the task $r_i$ to be selected, and $t_{r_i}$ is the time when $r_i$ was registered  and $\tau$ is the current time step. This is done between the lines 7 and 9 of Alg. 3. We select the tasks using this distribution for each chromosome initialization and we order them in the chromosome randomly (from line 10 to line 15). 

To create a new generation, we need to evaluate the quality of our previous population chromosomes thanks to a fitness function defined from our optimization objective. The goal is to minimize the result of this fitness function throughout the generations of the GA. The fitness function is defined as follows:
\begin{equation} \medmath{
    \alpha \Big ( \sum_{a \in A_\tau(H)} l_a \frac{1}{L_{max}} \Big ) +(1 -\alpha) \Big(1 - \frac{\sum_{a \in A_\tau(H)} \sum_{r \in R_\tau} x_{a,r}}{\lvert R_\tau \rvert} \Big)}
\end{equation}
\noindent where $l_a$ is the distance traveled by agent $a$ from its (current or anticipated) position in order to complete his new list of requests, and $r_n$ is the list of requests assigned to agent $a$ for this chromosome. We normalize the cost concerning  the total traveled distance agents $A_\tau(H)$ by $L_{max}$, which is the maximum distance traveled by all agents in the first generation of the GA. As we assign the selected tasks in a random order in the solution initialization, this maximum distance obtained in the first generation can be considered as a worst-case solution. The scoring of the different chromosomes is done between lines 20 and 22 in Alg. 3.

\vspace{-3mm}
\subsubsection{Evolution operations}
\vspace{-2mm}
Using the fitness function, we select the $30\%$ best chromosomes (line 24) and apply some operations on them to build a new generation. We use crossover operation and two types of mutation operations. These different operations are illustrated in Figure \ref{fig:operations}. 

The crossover operation selects two parent chromosomes and a random index. The elements from the first parent are copied until that index and from that index, we append the elements of the second parent, respecting the condition that a request can only be assigned to one agent. This creates a child chromosome (see line 29 of Alg. 3). After the crossover operation, some chromosomes go randomly through one of the mutations. This operation is usually used to avoid the genetic algorithm getting stuck on a local minimum (see \cite{Review_GA}). 
We call the first type of mutation swapping, which consists in taking two random tasks in a chromosome and swapping their positions. The other mutation is the inversion operation, where the order of the requests between two random indexes is inverted. A chromosome can (randomly) undergo a mutation following two probabilities $p_{muta}$ and $p_{swap}$ (line 30 to 34) which were empirically set.

\vspace{-3mm}
\subsubsection{Stopping condition}
\vspace{-2mm}
Our GA stops if it reaches one of two stopping conditions: (i) if it can not improve its solution value (e.g. not more than $\epsilon \simeq 0$) from the last couple of generations or (ii) if the time spent since the beginning of the GA calculations has reached the duration of the time window $\delta$ (line 18 in Alg. 3). 
If some requests were not assigned (e.g. due to the constraint on the number of tasks), they will be considered along with the buffer for the next allocation. The best solution in terms of fitness function evaluation is selected and we consider it as the assignment solution ($\mathrm{Sol}_\tau$) for the time window $\tau$ (line 37).


\vspace{-2mm}
\section{Experiments}
\label{sec:expes}
\vspace{-2mm}
\subsection{Benchmarks and Metrics}
\vspace{-2mm}

The first aspect we are interested in is how our proactive approach compares with a reactive approach that considers only the immediately available agents. For that, we are going to use two different benchmarks: the first one is a synthetic simulation and the second one is based on the real data of taxi dispatching in New York city\footnote{available at \url{http://www.andresmh.com/nyctaxitrips/}} obtained from reference \cite{Dickerson_2018}.

This is also the step where we have to make some adaptations to fit each scenario. We focused on the cost of the distance $l_a$ present. In our synthetic scenario, we consider the request as completed when the agent reaches it; so $l_a$ is only the distance from the anticipated position of the agent to  the first task and the distance between the other tasks. Whereas for taxi dispatching, we have to also take into account the pick-up and the arrival locations of the requests, so here $l_a$ is sum of the distance from the anticipated position of the taxi to the pick-up of the first request, the distance between the pick-up and the arrival for each task and the distance between the arrival location of the previous task and the pick-up of the following request. We can adapt this criterion to whatever scenario by just changing the formula of $l_a$.

During our synthetic simulation, we vary different parameters to analyze how they impact the performance of our proposal. We will first analyze the impact of changing $\alpha$ through empirical testing. Note that different values of $\alpha$ give more or less importance to either the traveled distance criterion or the percentage of assigned requests criterion. The goal is to find an $\alpha$ value that allows for a compromise between the performance of the different criteria. We will also evaluate how our method performs depending on the number of tasks compared to the number of agents: in one case, at each time window $\tau$ there will be fewer tasks to assign than the total number of agents $ \lvert R_\tau \rvert < \lvert A \rvert, \forall \tau \in \{0, ..., T-1\}$, and on the other hand, there will be more tasks than agents at each time window $\lvert R_\tau \rvert > \lvert A \rvert, \forall \tau \in \{0, ..., T-1\}$. 

In a second moment, we analyze the impact of the receding-horizon $H$. We denote by $H(k)$ the receding-horizon with a duration equal to $k$ time steps. In variable case, denoted $H(v)$,  we compute the solutions for several receding-horizon sizes and take the best solution among them. 
Note that the $H(0)$ case is equivalent to the classical reactive approach whereas the others constitute the proactive approach. In the variable receding-horizon case $H(v)$ we search a solution for $H(0),\ H(1), \ldots, H(5)$ and chose the best solution among them regarding the score from the fitness function in the GA. The last parameter we evaluate is the impact of the maximum amount of requests $C_{\tau,a}$ that can be assigned to an agent in each time window. For one case, we fixed $C_{\tau,a} = \frac{1}{3} |R_\tau|$ and for the other case $C_{\tau,a} = \infty$ for all different agents.

For the real data set of taxi dispatching, we want to see how our method scales and performs in a real-data based environment, where there is more ground to be covered by the numerous agents and the number of requests really varying according the time and the activity of the clients.

In terms of metrics we uses to evaluate our model, we are interested in three specific metrics registered during the total time length on our simulations. The first one is the total distance traveled by the agents, the second one is the time agents stay idle, and the last one is the percentage of allocated tasks. We want the distance and the idle time to be low and the percentage of assigned tasks to be high.  
The idle time is when agents stay idle waiting for their new assignments after completing their tasks. Normally, with the availability anticipation, this only happens to an agent if the timing to complete all their requests is less than $\tau$.

\vspace{-2mm}
\subsection{Results}
\vspace{-2mm}
\subsubsection{Synthetic benchmark}
\vspace{-2mm}
In this simulation, we have fixed the velocity of each agent at 1 m/s. The generated requests at $\tau$ have a random position in a squared grid world of dimensions of $10m \times 10m$. One simulation has a total of 30 time windows and we use 10 different simulations. During one simulation, each agent can travel a maximum of 150m. The results presented are based on the average of our 10 simulations.

\vspace{-2mm}
\paragraph{Evaluation of the impact of the $\alpha$ parameter.} We attributed to $\alpha$ the following values: \{0, 0.25, 0.5, 0.75 ,1\} for the fixed and variable receding-horizons. We also look at the case where there are fewer requests per time step than agents and the contrary. Results are presented in Table \ref{tab:alpha_results} in Appendix. We can observe that, in general, when the $\alpha$ value increases, the traveled distance decreases, which is an expected result since $\alpha$ is the weight of the distance in the fitness function. When $\alpha =0$, no attention is given to the way the tasks are ordered and the algorithm does not search for the shortest path. This is also when the idle time is the lowest since the agents have the biggest distance to cover.  
  
With $\lvert R_\tau \rvert > \lvert A \rvert$, when $\alpha$ value increases, the percentage of allocated tasks rises even if the weight of the total number or assigned tasks criterion is less important in the fitness function. With $\lvert R_\tau \rvert < \lvert A \rvert$, percentage of allocated tasks is almost 100 \% until $\alpha =0.75$, when the minimization of total distance becomes the dominant criterion and the percentage drops slightly. In this case, the the agents are underutilized and should be able to complete all the requests. For this reason, we chose to exclude $\alpha = 1$. The value with the best results is $\alpha = 0.75$. This value will used in the deep analysis presented in the following and on the rest of our synthetic benchmark tests. 

\begin{figure*}[!h]
\begin{subfigure}[b]{0.32\linewidth}
\centering
\includegraphics[width=\linewidth,trim=0cm 1cm 0cm 0cm, clip]{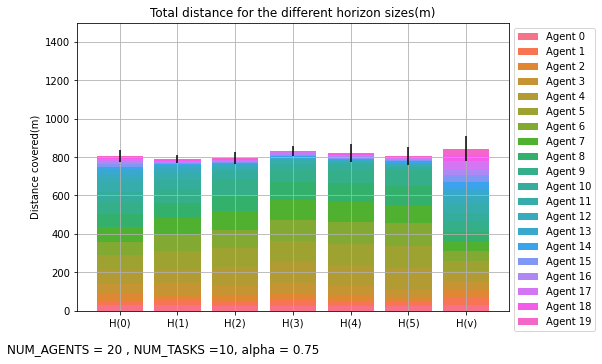} 
\caption{Total traveled distance}
\label{fig:subim1}
\end{subfigure}
\begin{subfigure}[b]{0.32\textwidth}
\centering
\includegraphics[width=\linewidth,trim=0cm 1cm 0cm 0cm, clip]{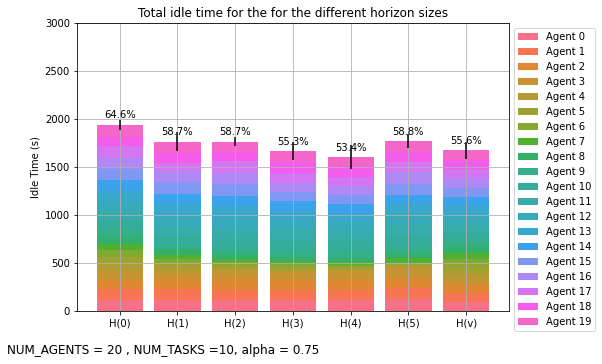}
\caption{Total idle time}
\label{fig:subim2}
\end{subfigure}
\begin{subfigure}[b]{0.32\textwidth}
\centering
\includegraphics[width=\linewidth,trim=0cm 1cm 0cm 0cm, clip]{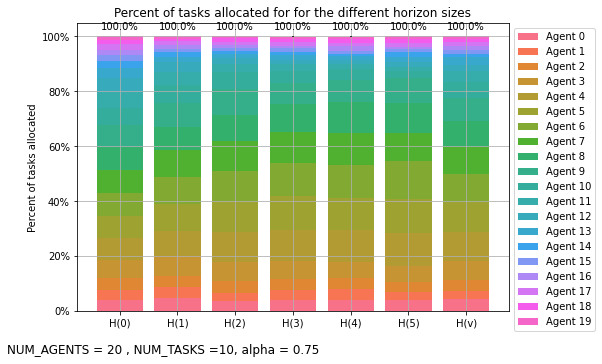}
\caption{Percent of assigned tasks}
\label{fig:subim2}
\end{subfigure}
\caption{Comparison between different receding-horizon sizes when $|R_\tau| < |A|$.}
\label{fig:lessagents}
\end{figure*}

\begin{figure*}[!h]
\begin{subfigure}{0.32\linewidth}
\centering
\includegraphics[width=\linewidth,trim=0cm 1cm 0cm 0cm, clip]{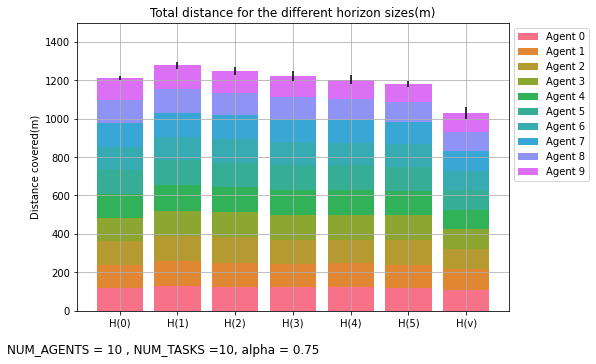} 
\caption{Total traveled distance}
\label{fig:subim1}
\end{subfigure}
\begin{subfigure}{0.32\linewidth}
\centering
\includegraphics[width=\linewidth,trim=0cm 1cm 0cm 0cm, clip]{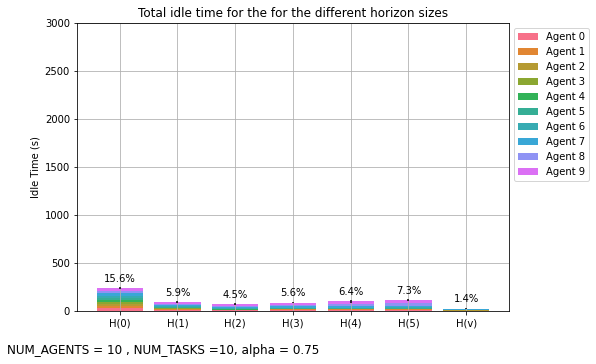}
\caption{Total idle time}
\label{fig:subim2}
\end{subfigure}
\begin{subfigure}{0.32\linewidth}
\centering
\includegraphics[width=\linewidth,trim=0cm 1cm 0cm 0cm, clip]{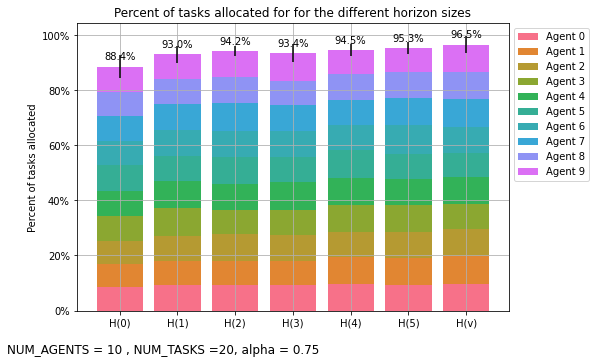}
\caption{Percent of assigned tasks}
\label{fig:subim2}
\end{subfigure}
\caption{Comparison for different receding-horizon sizes when $|R_\tau| >|A|$.}
\label{fig:moreagents}
\end{figure*}
\vspace{-2mm}
\paragraph{Less tasks than agents}
In Figure \ref{fig:lessagents}, we are paying attention to the case where at each time step the number of tasks is inferior to the number of agents. In particular, we have 20 agents and new 10 tasks at each time window $\tau$. In terms of total distance traveled during the simulation, we can see that it is almost equal for the different receding-horizon sizes.  For the total agents idle time, it is maximal when there is no anticipation (i.e. at $H(0)$) and decreases for the following values until it reaches a minimum in $H(4)$ before going slightly up again for $H(5)$. This decrease is explained by the fact that the availability anticipation allows to predict where and when the different agent finishes his tasks. The increase at the end can also be explained. The values of idle time are particularly high because the agents are underused. In terms of percentage tasks assigned, we can see that all of them have been assigned regardless of the size of the receding-horizon. In the case where there are less requests than agents, the anticipation availability method is not necessary as it does not bring advantages in terms of traveled distance and amount of allocated tasks (the two metrics we want to optimize) but can be interesting to lower the time agents stay idle.

\vspace{-2mm}
\paragraph{More tasks than agents}
Figure \ref{fig:moreagents} shows the results for the case where, at each round, the number of tasks is superior to the number of agents. Even in this case we maintain the constraint of $C_{\tau,a} = \frac{1}{3} |R_\tau|$. In general, we can see that with a farther receding-horizon, our method allows for a better balance. As previously, when there is no anticipation, the percentage of assigned tasks is the lowest and the agents idle time the highest. Here, with the constraint of maximum number of tasks per agent, we can see a significant improvement in terms of percentage of assigned tasks when anticipation is used.
With these results, we can clearly see that the variable receding-horizon is the best among all the receding-horizon cases, as it achieves the lowest distance, the lowest idle time and the higher percentage assigned tasks. This is due to the fact we take the receding-horizon size that is more advantageous, resulting in the best results. 
\vspace{-2mm}
\begin{figure*}[!h]
\begin{subfigure}{0.32\linewidth}
\centering
\includegraphics[width=\linewidth,trim=0cm 1cm 0cm 0cm, clip]{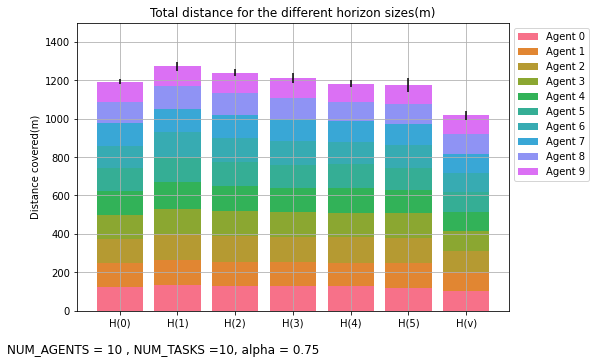} 
\caption{Total traveled distance}
\label{fig:subim1}
\end{subfigure}
\begin{subfigure}{0.32\linewidth}
\centering
\includegraphics[width=\linewidth,trim=0cm 1cm 0cm 0cm, clip]{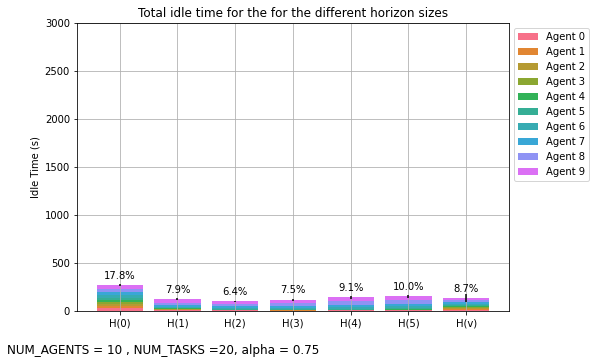}
\caption{Total idle time}
\label{fig:subim2}
\end{subfigure}
\begin{subfigure}{0.32\linewidth}
\centering
\includegraphics[width=\linewidth,trim=0cm 1cm 0cm 0cm, clip]{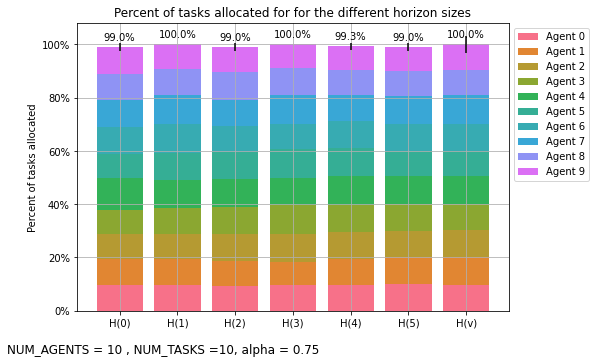}
\caption{Percent of assigned tasks}
\label{fig:subim2}
\end{subfigure}
\caption{Comparison for different receding-horizon sizes when $C_{\tau,a} =\infty$.}
\label{fig:relax}
\end{figure*}
\vspace{-2mm}
\begin{figure*}[!h]
\begin{subfigure}{0.32\linewidth}
\centering
\includegraphics[width=1.05\linewidth,trim=0cm 1cm 0cm 0cm, clip]{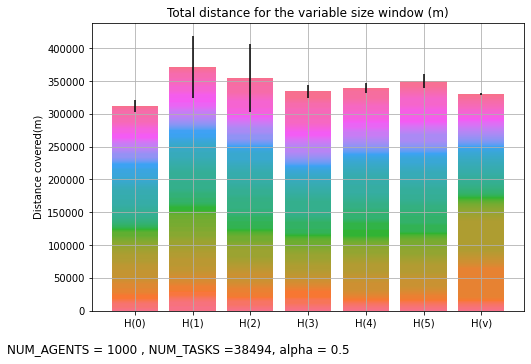} 
\caption{Total traveled distance}
\label{fig:subim1}
\end{subfigure}
\begin{subfigure}{0.32\linewidth}
\centering
\includegraphics[width=0.9\linewidth,trim=0cm 1cm 0cm 0cm, clip]{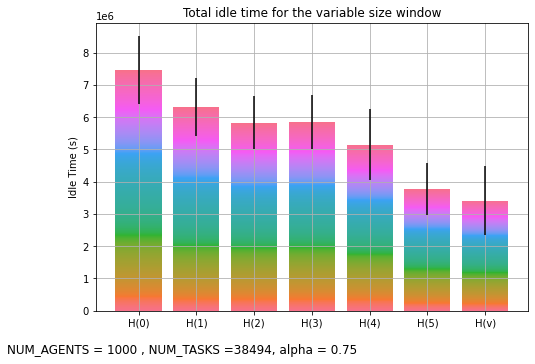}
\caption{Total drivers idle time}
\label{fig:subim2}
\end{subfigure}
\begin{subfigure}{0.32\linewidth}
\centering
\includegraphics[width=\linewidth,trim=0cm 1cm 0cm 0cm, clip]{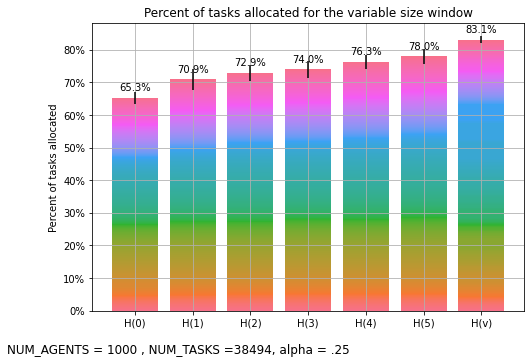}
\caption{Percent of assigned tasks}
\label{fig:subim2}
\end{subfigure}
\caption{Our method applied to taxi dispatching dataset to answer the consumers' requests.}
\label{fig:taxis}
\end{figure*}
\vspace{-2mm}
\paragraph{Relaxation of the $C_{\tau,a}$ constraint}

In this set of results we study the case where we relax the constraint $C_{\tau,a}$ such as $C_{\tau,a}=\infty$, with $\lvert R_\tau \rvert > \lvert A \rvert$ (we have 10 agents and 20 new tasks each time window $\tau$). Compared to the previous part, the total traveled distance is higher but it is explained by the fact that the agents are traveling to more tasks and at the same time. It is confirmed by the fact that the idle time is lower since our agents (e.g. resources) are more used. As before, the percentage of assigned requests is close to 100\%. 

For the different receding-horizon cases, in terms of total distance traveled, we have an increase from $H(0)$ to $H(1)$ and then the value decreases. The increase can be explained by the fact that the idle time with in these cases more important - i.e. less anticipation cases. With the receding-horizon size growing, the traveled distance diminishes because we balance the assigned tasks in a better way. Additionally, in terms of idle time, there is an important difference when there is no anticipation compared to when there is some anticipation, which allows for a better use of the agents. With the anticipation, there is no need to wait for the agents to finish their previous tasks before assigning them new ones. 

With these results we can speculate that our proactive online multi-assignment method which anticipates resource availability brings benefits in terms of resource use. 

\vspace{-3mm}
\subsubsection{Taxi Dispatching benchmark}
\vspace{-2mm}
We also applied our method to a real-set of data to analyze how our approach scales up. For this we use the dataset, which details the rides of the taxis in New York on the year 2013 (see \cite{Dickerson_2018}). To evaluate our method on this dataset we considered the pickup-time instead as the time a request for a ride was registered. We concentrated our test from January 07th to January 09th and during nights from 12:00am to 07:00am. We set a time window lasting 5 minutes. To answer the requests, we considered a constant fleet of 1000 taxis, randomly placed in New York. 
We assume the taxis to have a constant velocity of 30 mph (the authorized limit in New York). Contrarily to the synthetic benchmark, here the number of requests truly varies as it depends on the habits of the clients. This way, the minimum amount of requests are registered between 3AM and 4AM (most people are sleeping) whereas the maximum is close to 7AM when people start going to work.

The results are showed in Figure \ref{fig:taxis}. We can see that as in the synthetic benchmark simulations, the anticipation availability method allows for a higher percentage of assigned tasks, while also reducing the time the drivers stay idle. We also see that $H(v)$ is the setting with the more allocated tasks and one of the lowest distance, which highlights the potential of using the variable receding horizon. In a curious way, the distance does not steadily decrease like previously but this is probably due to the increase of the assigned requests since the distances are a lot more important in New York compared to our previous (synthetic) simulations.

There are also other practical advantages with our method. First of all, we see that our method is able to scale up well and keep with real-life demands. For comparison, the article \cite{Dickerson_2018} which also referenced this dataset used 30 tasks for allocating 550 tasks. With our method, we managed to allocate 1000 agents to around 30 000 tasks per night. We also get rid of human bias: in general, a driver working for a driving platform accepts or declines the request himself. Drivers may refuse some requests due to the distance, which leaves the consumer in the uncertainty if he will be picked up or not. Our method gets rid of this bias and tries to arrange the requests so that it can be combined with other close requests favoring ride-sharing. Moreover, we speculate that our method can also be used to determine fleet size to met demands, adding more agents when too many requests are being left unattended and reducing the number of agents if the idle time becomes important, leading to a variable size of fleet.%

\vspace{-5mm}
\section{Conclusion and Future Work}
\label{sec:conclusion}
\vspace{-2mm}
In order to treat the online multi-task assignment problem, we have proposed an alternative approach: online proactive multi-task assignment with resource availability anticipation, where we use a receding-horizon to anticipate which agents would be available with this horizon. Through our synthetic and real dataset benchmark simulations, we have shown that our method allows for generally assign a higher percentage of tasks to agents and that the agents tend to be less idle, leading to a better use of our resources. We have plans to further explore the concept of resource availability anticipation in the future. First of all, we plan to integrate a criterion to take into account the waiting times of the tasks for their completion, in order to try to minimize this waiting time. While the genetic algorithm is effective, it can be time-consuming in terms of calculation so we want to try to find another efficient but lighter allocation method. We are also planning to test our anticipation method with some algorithms that have been used for offline allocation or exploration but not in the context of online allocation. 

\nocite{*}
\bibliographystyle{eptcs}
\bibliography{generic}

\newpage
\section *{Appendix}
\label{sec:appendix}
\vspace{-4mm}
\begin{table} [htp]
  \begin{center}
    \caption{Average results for the evaluation of the impact of the $\alpha$ parameter given different receding-horizons and the cases where there are more/less agents than requests.}
    \label{tab:alpha_results}
    \scriptsize
    \begin{tabular}{*{8}{c}}
    \hline
    \rowcolor{gray!20}
    \multicolumn{8}{l}{\textbf{$\lvert R_\tau \rvert < \lvert A \rvert$}}\\
    \hline
    & \multicolumn{7}{c}{Total traveled distance (m)} \\
    \hline
    $\alpha$ & $H(0)$ & $H(1)$ & $H(2)$ & $H(3)$ & $H(4)$ & $H(5)$ & $H(v)$ \\ 
    \hline
    0 &1501 & 1524 & 1522 & 1501 & 1512 & 1487 & 1519 \\
    0.25 & 965 & 936 & 934 & 971 & 979 & 965 & 922 \\
    0.5 & 926 & 917 & 920 & 926 &  932 & 911 & 909 \\
    0.75 & 804 & 789 & 794 & 833& 821 & 805 & 843 \\
    1 & 867 & 839 & 874 & 886 & 886 & 864 & 859 \\
    \hline
    &\multicolumn{7}{c}{Idle time (s)} \\
    \hline
    0 & 1378.0 & 1350.0 & 1326.2 & 1330 & 1343.1 & 1350.9 & 1108.0\\
    0.25 & 1715.6 & 1572.6 & 1517.5 & 1500.0 & 1541.4 & 1536.4 & 1766.5 \\
    0.5 & 1736.7 & 1535.5 & 1585.2 & 1610.7 & 1528.4 & 1570.0 & 1733.2 \\
    0.75 & 1936.7 & 1759.6 & 1761.1 & 1659.0 & 1600.8 & 1765.4 & 1668.8 \\
    1 & 1860.1 & 1706.8 & 1681.8 & 1682.1 & 1646.1 & 1674.2 & 1608.3\\
    \hline
    & \multicolumn{7}{c}{Percentage of assigned tasks (\%)} \\
    \hline
    0 & 100 & 100 & 100 &100 & 100 & 100 & 100\\
    0.25 & 100 & 100 & 100 & 100 & 100 & 100 & 100\\
    0.5 & 100 & 100 & 100 & 100 & 100 & 100 & 100\\
    0.75 & 100 & 100 & 100 & 99.97 & 99.97 & 100 & 100 \\
    1 & 99.63 & 99.53 & 99.73 & 99.77 & 99.87 & 99.83 & 99.73 \\
    \hline
    \rowcolor{gray!20}
    \multicolumn{8}{l}{\textbf{$\lvert R_\tau \rvert > \lvert A \rvert$}} \\
    \hline
    & \multicolumn{7}{c}{Total traveled distance (m)} \\
    \hline
    $\alpha$ & $H(0)$ & $H(1)$ & $H(2)$ & $H(3)$ & $H(4)$ & $H(5)$ & $H(v)$ \\ 
    \hline
    0 & 1355 & 1421 & 1426 & 1437 & 1440 &1420 & 1273 \\
    0.25 & 1270 & 1334 &1315  & 1301 & 1274 & 1261 & 1138 \\
    0.5 &  1243 & 1297 & 1273 & 1250 & 1206 & 1202 & 1080 \\
    0.75 & 1211 & 1277 & 1247 & 1221  & 1203 & 1811 & 1031 \\
    1 & 1207 & 1277 & 1258 & 1225  & 1198 & 1208 & 1044 \\
    \hline
    &\multicolumn{7}{c}{Idle time (s)} \\
    \hline
    0 & 97.5 & 15.6 &7.3 & 9.8 &9.7 & 9.3 & 4.5 \\
    0.25 & 185.9 & 61.2 & 50.5 & 62.5 & 85.5 & 90.1 & 29.0 \\ 
    0.5 & 212.3 & 64.5 & 54.4 & 70.4 & 83.1 & 92.9 & 22.9 \\
    0.75 & 233.8 & 89.1 & 62.3 & 83.6 & 96.6 & 109.7 & 21.1 \\
    1 & 240.2 & 86.7 & 69.4 & 97.5 & 107.3 & 117.7 & 30.5\\
    \hline
    & \multicolumn{7}{c}{Percentage of assigned tasks (\%)} \\
    \hline
    0 & 79.5 & 86.4 &81.2 &84.3 &84.6 & 80.3 & 85.9 \\
    0.25 & 89.3 & 90.5 & 90.3 & 92.9 & 93.4 & 92.1 & 92.4 \\
    0.5 & 91.2 & 93.5 & 94.3 & 94.4 & 93.2 & 93.3 &95.9 \\
    0.75 & 88.8 & 93.0 & 94.2 & 93.4 & 94.5 & 95.3 & 96.5 \\
    1 &  89.5 & 93.1 & 94.4 & 93.7 & 93.4 & 94.7 & 97.9 \\
    \hline
    \end{tabular}
  \end{center}
\end{table}

\end{document}